\documentclass[pre]{revtex4}
\usepackage{graphicx}
\begin{document}
\title{Soliton lattices in the Gross-Pitaevskii equation with nonlocal and repulsive coupling}
\author{Hidetsugu Sakaguchi}
\affiliation{Department of Applied Science for Electronics and Materials,
Interdisciplinary Graduate School of Engineering Sciences, Kyushu
University, Kasuga, Fukuoka 816-8580, Japan}
\begin{abstract}
Spatially-periodic patterns are studied in nonlocally coupled Gross-Pitaevskii equation.  We show first that spatially periodic patterns appear in a model with the dipole-dipole interaction. Next, we study a model with a finite-range coupling, and show that a repulsively coupled system is closely related with an attractively coupled system and its soliton solution becomes a building block of the spatially-periodic structure.  That is, the spatially-periodic structure can be interpreted as a soliton lattice. An approximate form of the soliton is given by a variational method. Furthermore, the effects of the rotating harmonic potential and spin-orbit coupling are numerically studied. 
\end{abstract}
\maketitle
\section{Introduction}
Supersolid is an interesting material which is characterized with both the superfluidity and crystal structure~\cite{Andreev,Chester}. The superfluidity in solid $^4$He has been studied as an example of supersolid. Recently, the supersolid phases  are studied in Bose-Einstein condensates (BECs). The Gross-Pitaevskii equation is  a mathematical model of matter waves of BECs~\cite{Pethick,Stringari}. The GP equation is derived by a mean-field approximation of weakly interacting bosons at zero temperature. The effect of multi-body correlation and the fluctuation by finite temperature are neglected in the description by the GP equation. Various types of ground states  were successfully described by the GP equations. 
For example, a vortex lattice state was observed in rotating BECs with repulsive interaction~\cite{Williams,Matthews} and was described by the GP equation~\cite{Tsubota,Sakaguchi1}. Localized matter waves called solitons were found in experiments of BECs~\cite{Khaykovich} and the solitons were expressed as a ground state in the GP equation for Bose-Einstein condensate with attractive interaction~\cite{Malomed}. Pomeau and Rica studied a nonlocally coupled Gross-Pitaevskii (GP) equation as a model of supersolid~\cite{Pomeau}.
They showed the excitation spectrum exhibits the so-called roton minimum when a parameter exceeds a critical value which leads to the spatially periodic structure. Henkel, Nath, and Pohl studied the supersolid in Rydberg-dressed BECs~\cite{Henkel}. Since then, the supersolid in the Rydberg-dressed BECs has been studied by many authors~\cite{Maucher,Hsueh}. The GP equation with the finite-range interaction was also studied by several authors~\cite{Sepulveda,Kunimi,Ancilotto,Macri}. Prestipino et al. studied the spatially-periodic solution of the GP equation with a variational method~\cite{Prestipino}. Various other effects can be incorporated in the nonlocal GP equation. The effect of rotation to the Rydberg-dressed BECs was studied by Henkel et. al.~\cite{Henkel2}, and the effect of spin-orbit coupling was studied by Han et al.~\cite{Han}. Cinti showed the supersolid phase using a Monte-Carlo simulation of $N$ soft-core bosons~\cite{Cinti1,Cinti2}.

For a certain parameter sets, a strongly localized structure appears in the spatially periodic solution. In this paper, we would like to propose a viewpoint that the localized structure can be interpreted as a soliton to a GP equation with an attractive interaction, and the spatially periodic solution can be interpreted as a soliton lattice. We will show that the localized solution can be approximated as a soliton to the nonlinear Schr\"odinger equation at a certain parameter range in a one-dimensional system. 
In Sec.2, we show spatially periodic patterns in a model with dipole-dipole interaction which might be more easily realized in experiments, although similar results were already shown in the GP equation for the Rydberg-dresses BECS. In Secs.3 and 4, we study spatially-periodic patterns in a simpler finite-range interaction model. The model equation with the repulsive interaction is closely related with an attractively coupled system. A soliton appears naturally in the attractively coupled system. The soliton is a building block of the spatially-periodic structure in the repulsive system.  Furthermore, an approximate solution of the soliton is given by a variational method for the attractive GP equation. In Sec.5, spin-orbit coupled systems are studied, in which semi-vortex lattices are created. We show that the vortex structure is easily broken in a rotating harmonic potential. This type of spin-orbit coupled system under rotation has not been studied before. 
\section{Spatially periodic pattern in the one-dimensional Gross-Pitaevskii equations with dipole-dipole interaction}
The one-dimensional nonlocally coupled Gross-Pitaevskii equation is expressed as
\begin{equation}
i\frac{\partial \phi}{\partial t} =-\frac{\beta}{2}\nabla^2\phi+\left \{ \int G(x-x^{\prime})|\phi(x^{\prime})|^{2}dx^{\prime}\right \}\phi,
\end{equation}
where the Planck constant is set to be 1, $m_0=1/\beta$ denotes a parameter proportional to particle mass. The second term on the right-hand side represents the repulsive and nonlocal interaction, where $G(x-x^{\prime})$ is an integral kernel characterizing the nonlocal coupling. 
For the Rydberg-dressed BECs, the integral kernel takes a form $G(x-x^{\prime})=g/[\epsilon+(x-x^{\prime})^6/R^6]$. For the dipole-dipole interaction, $G(x-x^{\prime})=g/[\epsilon+|x-x^{\prime}|^3/R^3]$.  
There is a uniform solution to Eq.~(1): $\phi(x)=Ae^{-i\mu t}$ where $\mu=A^2\int G(x)dx$. The modulational instability of the uniform solution can be investigated by assuming the wave function as $\phi=A(1+\delta(x))e^{i\mu t+i\theta(x)+\lambda t}$. The eigenvalue $\lambda(k)$ for the perturbation with wavevector $k$ is given by 
\begin{equation}
\lambda^2(k)=-\frac{\beta k^2}{2}\left (\frac{\beta k^2}{2}+G(k)A^2\right ),
\end{equation}
where  $G(k)=\int G(x)e^{ikx}dx$.
If $\lambda^2(k)<0$, $\lambda$ is pure imaginary and the uniform state is stable against the perturbation, however, if $\lambda^2(k)>0$, $\lambda$ takes a real positive value and the uniform state becomes unstable. This corresponds to the roton softening in the excitation spectrum~\cite{Pomeau,Macri}.

As an example, we show the instability in the case of the dipole-dipole interaction, since the nonlocal GP equation has been mainly studied for the Rydberg-dressed BECs and the finite-range interaction. Figure 1(a) shows $\lambda^2(k)$ for $\beta=0.1$ and the dipole-dipole interaction with $g=1$, $\epsilon=1$, and $R=5$.  The eigenvalue becomes positive for $k\sim 1$. Figure 1(b) shows a ground state of Eq.~(2) in one dimension obtained by the imaginary-time simulation method. In the imaginary-time evolution,  the numerical simulation of the equation obtained by replacing $t$ with $-i\tau$ from Eq.~(1) is performed under the condition of the constant norm $N=\int_0^L|\phi|^2dx$. Generally, the ground state is obtained by the imaginary-time evolution method. The total norm  $N$ is fixed to be $30$. The system size $L$ is 30 and the periodic boundary conditions are imposed at the boundaries $x=-L/2$ and $L/2$. A periodic pattern of wavenumber $L/5=6$ appears as a result of the instability. Figure 1(c) shows the real time evolution of $|\phi|$ by Eq.~(1) after $t=1000$ starting from the initial condition $\phi(x)=1+r(x)$ where $r(x)$ is a random number between -0.1 and 0.1. A spatially periodic pattern with five peaks appears from the randomly perturbed uniform initial condition even in the real time evolution of Eq.~(1).  
\begin{figure}[h]
\begin{center}
\includegraphics[height=4.cm]{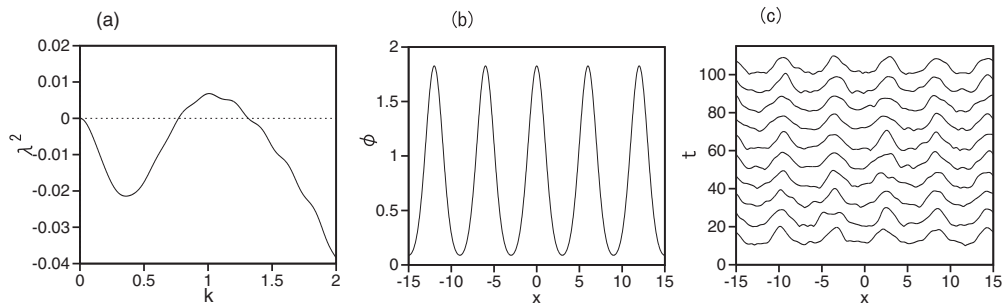}
\end{center}
\caption{(a) $\lambda^2(k)$ for the dipole-dipole interaction with $g=1$, $\epsilon=1$, $R=5$, and $\beta=0.1$. (b) The ground state obtained by the imaginary-time evolution method. (c) Real time evolution of Eq.~(1) starting from the initial condition $\phi(x)=1+r(x)$ where $r(x)$ is a random number between -0.1 and 0.1.}
\label{fig1}
\end{figure}
\section{Soliton lattices in the Gross-Pitaevskii equation with finite-range coupling}
Hereafter, we consider a simple finite-range coupling expressed as $G(x-x^{\prime})=g$ for $|x-x^{\prime}|\le L_g$ and 0 for $|x-x^{\prime}|> L_g$.  This function can be expressed as a limit of $g/(1+|x-x^{\prime}|^n/L_g^n)$ including both the dipole-dipole interaction and the interaction in the Rydberg dressed BECs, i.e., $G(x-x^{\prime})=\lim_{n\rightarrow\infty}g/[1+|x-x^{\prime}|^n/L_g^n]$. Furthermore, the parameter $\beta$ is set to 1. The model equation is rewritten as \begin{equation}
i\frac{\partial \phi}{\partial t} =-\frac{1}{2}\frac{\partial^2\phi}{\partial x^2}+g\left \{\int_{x-L_g}^{x+L_g}|\phi(x^{\prime})|^{2}dx^{\prime}\right \}\phi.
\end{equation}
Figure 2(a) shows a spatially-periodic structure obtained by the imaginary-time evolution method at $g=5$, $L=20$, $L_g=5.5$, and $N=1$. Rhombi in Fig.~2(b) show the peak amplitudes of periodic solutions as a function of $L_g$ for $g=5,L=20$, and $N=1$. Spatially-periodic solutions of wavelength $L/3$ appears at $L_g=3.8$ when $L_g$ increases.  The periodic solution of mode $m=3$ changes to a solution of mode $m=2$ at $L_g=5.6$, and the one-hump solution of $m=1$ appears at $L_g=10.1$. 
  
Equation (3) can be rewritten as
\begin{eqnarray}
i\frac{\partial \phi}{\partial t} &=&-\frac{1}{2}\frac{\partial^2\phi}{\partial x^2}+g\left \{\int_{x-L}^{x+L}|\phi(x^{\prime})|^{2}dx^{\prime}-\int_{x-L}^{x-L_g}|\phi(x^{\prime})|^{2}dx^{\prime}-\int_{x+L_g}^{x+L}|\phi(x^{\prime})|^{2}dx^{\prime}\right \}\phi\nonumber\\
&=&-\frac{1}{2}\frac{\partial^2\phi}{\partial x^2}+2gN\phi-g\left \{\int_{x-(L-L_g)}^{x+(L-L_g)}|\phi(x^{\prime})|^{2}dx^{\prime}\right \}\phi,
\end{eqnarray}
where the total norm is $N=\int_0^L|\phi|^2dx$ and the spatial periodicity $\phi(x+L)=\phi(x-L)=\phi(x)$ is used.
Note that the nonlocal repulsive interaction is changed into an attractive interaction. The BECs tend to be localized owing to the attractive interaction. 
This is a reason why a soliton-like localized structure appears.  The soliton-like structure for mode $m$ can be expressed as a solution to  
\begin{equation}
i\frac{\partial \phi}{\partial t}=-\frac{1}{2}\frac{\partial^2\phi}{\partial x^2}-g\left \{\int_{x-(L^{\prime}-L_g)}^{x+(L^{\prime}-L_g)}|\phi(x^{\prime})|^{2}dx^{\prime}\right \}\phi,\end{equation}
in a system of size $L^{\prime}=L/m$, and the norm is $N^{\prime}=N/m$.  Here, the term $2gN\phi$ in Eq.~(4) is neglected because the term can be removed by the transformation $\phi e^{-i 2gNt}\rightarrow \phi$, and the term is irrelevant to the spatial structure of the solution. The solid line in Fig.~2(b) shows the peak amplitude of the one-hump solution to the nonlocally-coupled attractive model equation Eq.~(5) for $m=1,2,$, and 3. Good agreement is seen between the numerical results (rhombi) by the direct numerical simulations of Eq.~(3) and those for Eq.~(5) (solid lines).      

The spatially periodic solution to Eq.~(3) can be expressed as $m$ trains of  one-hump solutions to Eq.~(5). 
When $L_g$ is changed as a control parameter, the peak amplitude of the soliton takes the maximum at a value slightly smaller than $L/m$ and decreases rapidly to zero near $L/m$ for each $m$. A strongly localized structure appears near the maximum point. We call the localized structure a soliton in this paper. The soliton generally appears in the GP equation with attractive interaction~\cite{Malomed}. The soliton appears because the originally repulsive GP equation (3) can be transformed to an attractive GP equation (5). 

If $(L-L_g)$ is sufficiently small, Eq.~(5) for $m=1$ is approximated at 
\begin{equation}
i\frac{\partial \phi}{\partial t}=-\frac{1}{2}\frac{\partial^2\phi}{\partial x^2}-2g(L-L_g)|\phi|^2\phi.
\end{equation}
This is equivalent to the nonlinear Schr\"odinger equation. The soliton solution to Eq.~(6) is expressed as
\begin{equation}
\phi=\frac{A}{\cosh Bx},
\end{equation}
where $2A^2/B=N$ and $2g(L-L_g)A^2=B^2$.
The solid line in Fig.~2(c) shows one-hump solution to Eq.~(5) for $m=1$ at $L_g=19.6$. The dashed line in Fig.~2(c) denotes the soliton solution $1/\cosh(2x)$ to Eq.~(6). The pulse solution is well approximated by the soliton solution. 
The solid line in  Fig.~2(d) shows the peak amplitude for the one-hump solution of Eq.~(5) for $m=1$ as a function of $L_g$. The dotted line denotes $A=\sqrt{2.5(L-L_g)}$ for the soliton solution. Good agreement is observed for $L_g>19.4$. However, the peak amplitude increases with $L_g$ for $L_g<19.4$. 

The total energy $E$ of the solution to Eq.~(5) for $m=1$ is expressed as
\begin{equation}
E=\frac{1}{2}\int \left |\frac{\partial \phi}{\partial x}\right |^2dx-\frac{g}{2}\int_0^L\int_{x-(L-L_g)}^{x+(L-L_g)}|\phi(x)|^2|\phi(x^{\prime})|^2dxdx^{\prime}.
\end{equation}
If the Gaussian function is assumed for $\phi$, i.e., $\phi=Ae^{-\alpha x^2}$, $E$ is expressed as 
\begin{equation}
E(\alpha)=\frac{N}{2}\alpha-\frac{2}{\sqrt{\pi}}gN^2\int_0^{\sqrt{2\alpha}L_g}e^{-z^2}dz,
\end{equation}
where $N=\sqrt{\pi/(2\alpha)}A^2$. The variational approximation gives $\alpha$ from the minimization of $E(\alpha)$ and $A$ is obtained from $A=(N\sqrt{2\alpha/\pi})^{1/2}$. The dashed line in Fig.~2(d) shows the relationship between $A$ and $L_g$ obtained by the variational method, which is a good approximation for $L_g<18.2$.  
\begin{figure}[h]
\begin{center}
\includegraphics[height=3.5cm]{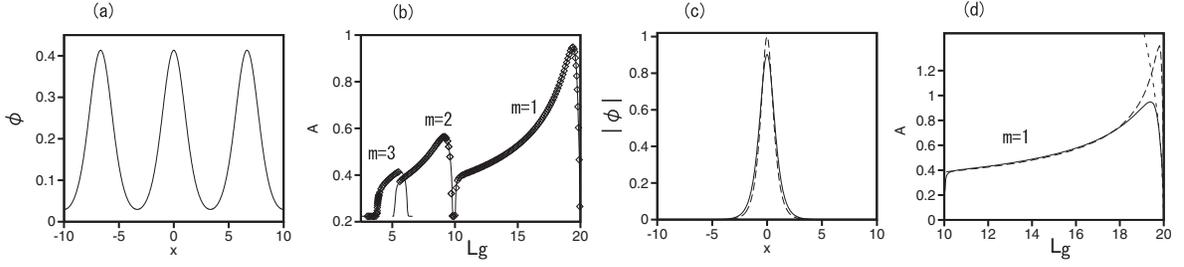}
\end{center}
\caption{(a) Periodic structure of wavelength $L/3$ at $g=5$, $L=20$, $L_g=5.5$, and $N=1$. (b) Peak amplitude $A$ as a function of $L_g$. Rhombi denote numerical results of Eq.~(3) and solid lines denote numerical results of Eq.~(5) for $m=1,2$, and 3. (c) Pulse solution (solid line) at $g=5$, $L=20$, $L_g=19.6$, and $N=1$, and the soliton solution (dashed line). (d) Comparison of numerically obtained peak amplitudes (solid line) with the variational approximation (dashed line) and the soliton solution (dotted line).}
\label{fig2}
\end{figure}

\section{Two-dimensional Gross-Pitaevskii equation} 
Next, we study the two dimensional nonlocally coupled GP equation. The two-dimensional integral kernel is expressed as  $G(x-x^{\prime},y-y^{\prime})=g$ only for $|x-x^{\prime}|\le L_g$ and $|y-y^{\prime}|\le L_g$. The model equation is written as
\begin{equation}
i\frac{\partial \phi}{\partial t}=-\frac{1}{2}\nabla^2\phi+g\left \{\int\int_S|\phi(x^{\prime},y^{\prime})|^{2}dx^{\prime}dy^{\prime}\right \}\phi,
\end{equation} 
where the integral region $S$ satisfies  $|x-x^{\prime}|\le L_g$ and $|y-y^{\prime}|\le L_g$. Figure 3(a) shows a spatially periodic stationary solution of wavelength $L/3$ at $g=20$, $L=20$, $L_g=5$, and $N=16$. The region satisfying $|\phi|>0.3$ is plotted in this figure. The rhombi in Fig.~3(b) shows the peak amplitude as a function of $L_g$ for  $g=20$, $L=20$, and $N=16$. Spatially-periodic states with wavelength $L/m$ ($m=1,2,3$, and 4) appears by changing $L_g$. 
The solid line shows the peak amplitude of the corresponding attractive model equations in a system of $(L/m)\times (L/m)$ for $m=1,2,3,$ and 4:
\begin{equation}
i\frac{\partial \phi}{\partial t}=-\frac{1}{2}\nabla^2\phi-g\left \{\int\int_{S^{\prime}}|\phi(x^{\prime},y^{\prime})|^{2}dx^{\prime}dy^{\prime}\right \}\phi,
\end{equation} 
where $S^{\prime}$ is a region satisfying $|x-x^{\prime}|>L_g/m$ or $|y-y^{\prime}|>L_g/m$, and the norm is $N^{\prime}=N/m^2$.  
Good agreement is seen in numerical results for Eqs.~(10) and (11), although discontinuous mode transitions occur in the direct numerical simulation of Eq.~(10). Intermediate states sometimes appear near the transition of modes. An example of the intermediate state between $m=3$ and 4 at $L_g=4.25$ is shown in Fig.~3(c).  

The total energy $E$ of Eq.~(11) for $m=1$ is expressed as
\begin{equation}
E=\frac{1}{2}\int \left |\nabla\phi\right |^2dx-\frac{1}{2}\int_0^L\int_0^L \int\int_{S^{\prime}} |\phi(x,y)|^2|\phi(x^{\prime},y^{\prime})|^2dx^{\prime}dy^{\prime}dxdy.
\end{equation}
If the Gaussian function is assumed for $\phi$, i.e., $\phi=Ae^{-\alpha (x^2+y^2)}$, $E$ is expressed as 
\begin{equation}
E(\alpha)=N\alpha-\frac{2}{\pi}gN^2\{I_S^2+2(0.5-I_S)I_S\},
\end{equation}
where $I_S=\int_0^{\sqrt{\alpha}L_g}e^{-z^2}dz$, and $N=\pi A^2/(2\alpha)$. The variational approximation gives $\alpha$ from the minimization of $E(\alpha)$ and $A$ is obtained from $A=(2N\alpha/\pi)^{1/2}$. The solid line in Fig.~3(d) is direct numerical results of Eq.~(11) for $m=1$. The dashed line in Fig.~3(d) shows the relationship between $L_g$ and $A$ obtained by the variational method, which is a good approximation for $L_g<18$. The peak amplitude takes the maximum value at a certain $L_g$ slightly smaller than $L/m$, and the peak amplitude decreases rapidly and becomes 0 near $L/m$. That is, a very strongly localized structure or a two-dimensional soliton of large amplitude appears near $L_g=L/m$. The behavior is similar to the one-dimensional system, however, the change of the peak amplitude near $L/m$ is even larger in the two-dimensional system. 
However, there is no collapse phenomenon in the nonlocally coupled equation Eq.~(11) with attractive interaction in contrast to the locally coupled two-dimensional nonlinear Schr\"odinger equation.   
\begin{figure}[h]
\begin{center}
\includegraphics[height=4.cm]{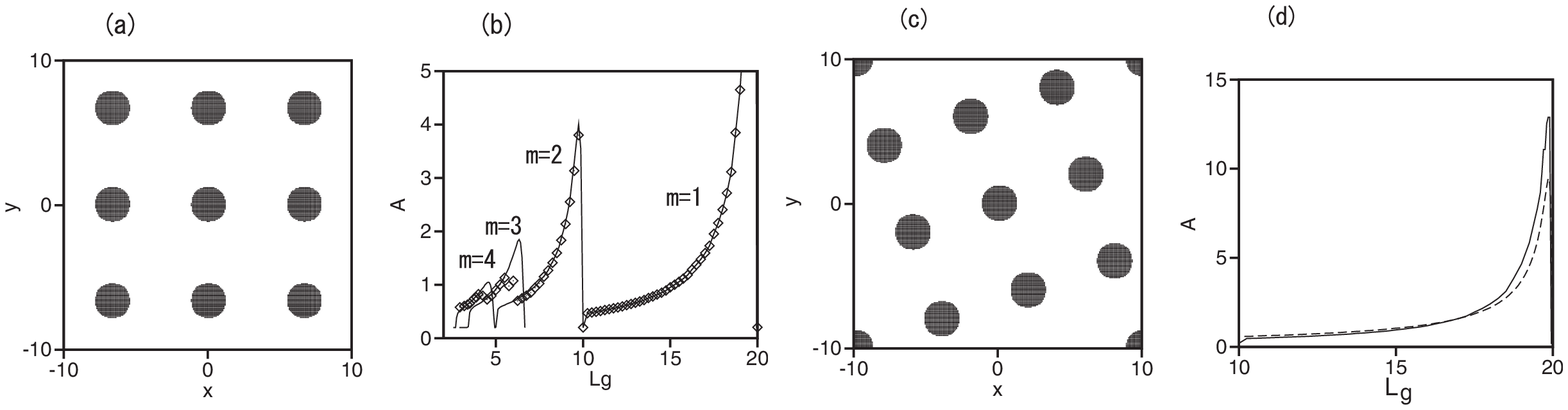}
\end{center}
\caption{(a) Periodic structure of period $L/3$ at  $g=20$, $L=20$, $L_g=5$, and $N=16$. (b) Peak amplitude $A$ as a function of $L_g$. Rhombi show numerical results of Eq.~(10) and solid lines denotes numerical results of Eq.~(11). 
(c) Intermediate state between $m=3$ and 4 at  $g=20$, $L=20$, $L_g=4.25$, and $N=16$. (d) Comparison of numerically obtained peak amplitudes of Eq.~(11) for $m=1$ (solid line) with the variational approximation (dashed line).}
\label{fig3}
\end{figure}

If the BECs are trapped in a rotating harmonic potential, the GP equation is expressed with
\begin{equation}
i\frac{\partial \phi}{\partial t} =-\frac{1}{2}\nabla^2\phi+g\left\{\int \int_S|\phi(x^{\prime},y^{\prime})|^{2}dx^{\prime}dy^{\prime}\right\}\phi+\frac{1}{2}K(x^2+y^2)\phi-\Omega L_z\phi, 
\end{equation}
where $L_z=-i(x\partial/\partial y-y\partial/\partial x)$ and $\Omega$ is the frequency of rotation. When $K=\Omega^2$, Eq.~(14) becomes
\begin{equation}
i\frac{\partial \phi}{\partial t} =-\left [\frac{1}{2}\left (i\frac{\partial}{\partial x}-\Omega y\right )^2+\frac{1}{2}\left (i\frac{\partial}{\partial y}+\Omega x\right )^2\right ]\phi+g\left\{\int\int_S|\phi(x^{\prime},y^{\prime})|^{2}dx^{\prime}dy^{\prime}\right\}\phi. 
\end{equation}
If $g=0$, this equation is equivalent to the 2D Schr\"odinger equation of a charged particle in uniform magnetic field $\Omega$ directed perpendicular to the $(x,y)$ plane. For the locally coupled Gross-Pitaevskii equation with $g>0$, a vortex lattice state appears~\cite{Williams,Matthews,Tsubota,Sakaguchi1}. Figure 4(a) shows a spatially-periodic state of wavelength $L/4$  at $\Omega=1$, $g=20$, $L=20$, $L_g=3.5$, and $N=16$. In Fig.~4(a), $|\phi|>0.3$ is satisfied in the green region and Re$\phi>0.3$ is satisfied in the blue region. The wave function $\phi$ changes in the azimuthal direction like $\phi(x,y)\sim e^{ik(r)\theta}$ where $\theta$ is the angle around the center $(0,0)$.  Figure 4(b) shows the numerically obtained wavenumber $k(r)$ at several positions near the peaks of $|\phi|$  as a function of the distance from the center at $\Omega=1$. The dashed lines denote $k(r)=\Omega r$. The wavenumber $k(r)$ increases with $r$ and the proportionality coefficient is equal to $\Omega$. The term of $-\Omega L_z$ in Eq.~(15) induces the clockwise rotation of frequency $\Omega$, which is compensated by the anti-clockwise rotation induced by the phase factor $e^{ik(r)\theta}$, and a stationary state appears.   
We can also interpret that this state is a vortex lattice state where vortices are expelled from the regions of the soliton lattice to the region of $|\phi|\sim 0$ and only the phase factor survives. 
\begin{figure}[h]
\begin{center}
\includegraphics[height=4.5cm]{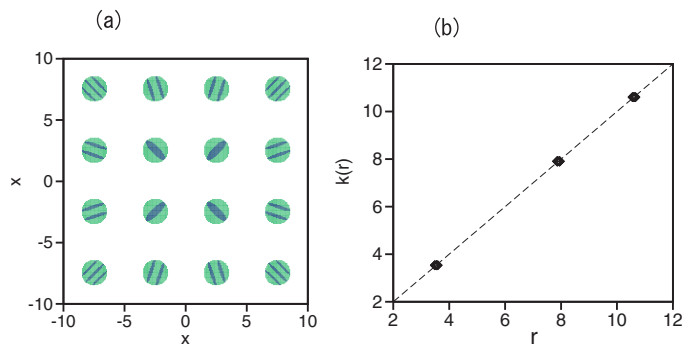}
\end{center}
\caption{(a) Periodic structure of period $L/4$ at $\Omega=1$, $g=20$, $L=20$, $L_g=3.5$, and $N=16$. (b) Wavenumber $k(r)$ as a function of $r$ at $\Omega=1$. The dashed line is $k(r)=r$.}
\label{fig4}
\end{figure}

\section{Spin-orbit coupled Bose-Einstein condensates}
Spin-orbit coupled BECs have been intensively since Lin et al. realized the spin-orbit coupling in BECs experimentally~\cite{Lin}. In the spin-orbit coupled BECs, the spin-up and spin-down components interact with each other through the spin-orbit coupling $k_y\sigma_x-k_x\sigma_y$ where $k_x$ and $k_y$ are the momentum operator in the $x$ and $y$ directions and $\sigma_{x,y}$ are the Pauli matrix. The spin-orbit coupled GP equations have been studied by several authors~\cite{Zhai,Malomed2}. Many remarkable nonlinear phenomena such as stripe patterns and solitons have been found in the spin-orbit coupled GP equations. Sakaguchi et al. found a vortex soliton called a semi-vortex in two-dimensional spin-orbit coupled GP equation~\cite{Sakaguchi2}. In the semi-vortex, the spin-up component takes a form of a two-dimensional soliton and the spin-down component takes a form of localized vortex.   
 A nonlocally coupled system with spin-orbit interaction in two dimensions is expressed as
\begin{eqnarray}
i\frac{\partial \phi_{+}}{\partial t} &=&-\frac{1}{2}\nabla^2\phi_{+}+g\left \{\int \int_S(|\phi_{+}|^{2}+|\phi_{-}|^2)dx^{\prime}dy^{\prime}\right \}\phi_{+}+\lambda\left (\frac{\partial \phi_{-}}{\partial x}-i\frac{\partial\phi_{-}}{\partial y}\right ),\nonumber \\
i\frac{\partial \phi_{-}}{\partial t} &=&-\frac{1}{2}\nabla^2\phi_{-}+g\left \{\int\int_S(|\phi_{-}|^2+|\phi _{+}|^{2})dx^{\prime}dy^{\prime}\right \}\phi_{-}-\lambda\left (\frac{\partial \phi_{+}}{\partial x}+i\frac{\partial \phi_{+}}{\partial y}\right ),\nonumber\\ \label{2dso}
\end{eqnarray}
where $\lambda$ is the strength of the spin-orbit coupling and $S$ is the integral region satisfying $|x-x^{\prime}|<L_g$ and $|y-y^{\prime}|<L_g$.

Figures 5(a) and (b) show the regions satisfying $|\phi_+|>0.2$ and $|\phi_{-}|>0.2$ for $g=20$, $L=20$, $L_g=5$, $N=16$, and $\lambda=1$. A lattice of semi-vortices with wavelength $L/3$ appears~\cite{Han}. A vortex exists at each center of annular structure of $|\phi_{-}|$. Similarly to the previous section, each semi-vortex can be expressed by the attractively interacting model equation:
\begin{eqnarray}
i\frac{\partial \phi_{+}}{\partial t} &=&-\frac{1}{2}\nabla^2\phi_{+}-g\left \{\int\int_{S^{\prime}} (|\phi_{+}|^{2}+|\phi_{-}|^2)dx^{\prime}dy^{\prime}\right \}\phi_{+}+\lambda\left (\frac{\partial \phi_{-}}{\partial x}-i\frac{\partial\phi_{-}}{\partial y}\right ),\nonumber\\
i\frac{\partial \phi_{-}}{\partial t} &=&-\frac{1}{2}\nabla^2\phi_{-}-g\left \{\int\int_{S^{\prime}}(|\phi_{-}|^2+|\phi _{+}|^{2})dx^{\prime}dy^{\prime}\right \}\phi_{-}-\lambda\left (\frac{\partial \phi_{+}}{\partial x}+i\frac{\partial \phi_{+}}{\partial y}\right ),\nonumber\\ \label{2dso2}
\end{eqnarray}
where $S^{\prime}$ is a region satisfying $|x-x^{\prime}|>L_g/m$ or $|y-y^{\prime}|>L_g/m$. 
Figure 5(c) compares the peak amplitudes $A$ of $|\phi_{+}|$ of the semi-vortex lattice in Eq.~(16) with the peak amplitudes  of $|\phi_{+}|$  obtained by Eq.~(17). Good agreement is seen. It is noted that discontinuous transition from $m=3$ to $m=2$ modes occurs near $L_g=6.2$ in Eq.~(16). These results are similar to the ones shown in Fig.~4. 
\begin{figure}[h]
\begin{center}
\includegraphics[height=4.cm]{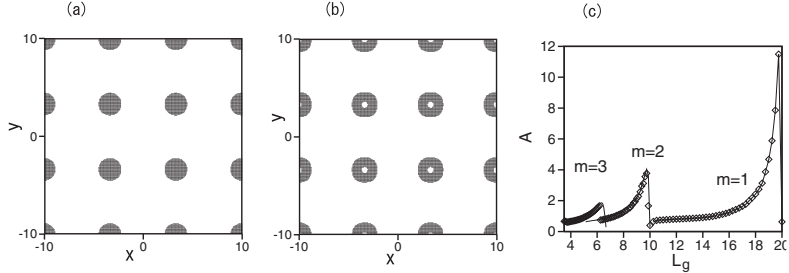}
\end{center}
\caption{Periodic structure of (a) $|\phi_{+}|$ and (b) $|\phi_{-}|$ with wavelength $L/3$ at  $g=20$, $L=20$, $L_g=5$, $\lambda=1$, and $N=16$. Regions satisfying $|\phi_{\pm}|>0.2$ are plotted. (c) Peak amplitude $A$ as a function of $L_g$. Rhombi show numerical results of Eq.~(16) and solid lines denote numerical results of Eq.~(17).}
\label{fig5}
\end{figure}

If the spin-orbit coupled BECs are trapped in the rotating harmonic potential satisfying $K=\Omega^2$, the coupled GP equations are expressed as~\cite{Sakaguchi3}
\begin{eqnarray}
i\frac{\partial \phi_{+}}{\partial t} &=&-\left [\frac{1}{2}\left (i\frac{\partial}{\partial x}-\Omega y\right )^2+\frac{1}{2}\left (i\frac{\partial}{\partial y}+\Omega x\right )^2\right ]\phi_{+}\nonumber\\
& &+g\left \{\int\int_{S} (|\phi_{+}|^{2}+|\phi_{-}|^2)dx^{\prime}dy^{\prime}\right \}\phi_{+}+\lambda\left (\frac{\partial \phi_{-}}{\partial x}-i\frac{\partial\phi_{-}}{\partial y}\right ),\nonumber \\
i\frac{\partial \phi_{-}}{\partial t} &=&-\left [\frac{1}{2}\left (i\frac{\partial}{\partial x}-\Omega y\right )^2+\frac{1}{2}\left (i\frac{\partial}{\partial y}+\Omega x\right )^2\right ]\phi_{-}\nonumber\\
& &+g\left \{\int\int_{S}(|\phi_{-}|^2+|\phi _{+}|^{2})dx^{\prime}dy^{\prime}\right \}\phi_{-}-\lambda\left (\frac{\partial \phi_{+}}{\partial x}+i\frac{\partial \phi_{+}}{\partial y}\right ). \label{2rdso}
\end{eqnarray}
The semi-vortex lattice disappears when $\Omega$ increases.
Figures 6(a)-(c) show regions satisfying $|\phi_{-}|>0.1$ at (a) $\Omega=0.001$, (b) 0.002, and (c) 0.004 for $L=20$, $g=20$, $L_g=3.5$, $\lambda=1$, and $N=16$. 
Spatially-periodic semi-vortex lattices of wavelength $L/4$ appear at $L_g=3.5$. When $\Omega=0$, the vortices of $\phi_{-}$ locate at the center of each localized structure. The vortex positions move toward the origin $(0,0)$ of the whole system. The movement is larger for the semi-vortices which is more distant from the origin. At $\Omega=0.001$, the vortices near the four corners go out of the shaded regions representing the localized structure. At $\Omega=0.002$, only the four vortices near the center survive.
 At $\Omega=0.004$, all vortices go out from the shaded regions. That is, the semi-vortex structure is broken at relatively small $\Omega$. The peak amplitude of $|\phi_{+}|$ and $|\phi_{-}|$ take a similar value at $\Omega=0.02$.    
\begin{figure}[h]
\begin{center}
\includegraphics[height=4.cm]{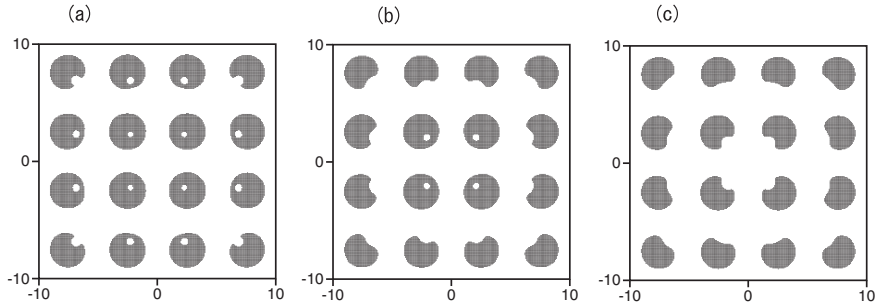}
\end{center}
\caption{Vortex lattices with wavelength $L/4$ of $\phi_{-}$ at (a) $\Omega=0.001$, (b) 0.002, and (c) 0.004 at  $g=20$, $L=20$, $L_g=3.5$, $\lambda=1$, and $N=16$. Regions satisfying $|\phi_{-}|>0.1$ are plotted. 
.}
\label{fig6}
\end{figure}
\section{Summary} 
First, we have shown that a spatially-periodic structure can be created in a system with the dipole-dipole type nonlocal coupling. BECs with the dipole-dipole interaction might be more easily realized than the Rydberg-dressed Bose-Einstein condensates studied previously. Next, we have studied a model system with a finite-range nonlocal interaction more in detail. A spatially-periodic pattern like a square lattice appears in the repulsively coupled system in two dimensions owing to the anisotropy. Dynamical transitions between spatially periodic structures with different wavenumber occur when the interaction range $L_g$ is changed. 
We have found that the peak amplitude takes a sharp maximum slightly below $L_g=L/m$ ($m$ integer) when $L_g$ is changed as a control parameter. The spatially periodic structure can be interpreted as a soliton lattice, in that the soliton structure is expressed by a solution to an attractively coupled system and works as a building block of the spatially-periodic pattern. Approximate solutions were derived by the variational method. In a rotating harmonic potential, a phase wave is induced in the azimuthal direction as $\phi\sim e^{ik(r)\theta}$.  The wavevector along the azimuthal direction increases in proportion to the distance from the center of the harmonic potential, which implies the rigid rotation. In the spin-orbit coupled nonlocal model, semi-vortex lattices are created, however, the vortex structure is easily broken in the rotating harmonic potential.      
\end{document}